\documentclass[aps,prd]{revtex4}
\usepackage{amsfonts}
\newcommand{\nn}{\nonumber}
\newcommand{\be}{\begin{equation}}
\newcommand{\ee}{\end{equation}}
\newcommand{\bd}{\begin{displaymath}}
\newcommand{\ed}{\end{displaymath}}
\newcommand{\bea}{\begin{eqnarray}}
\newcommand{\eea}{\end{eqnarray}}
\renewcommand{\paragraph}[1]{
\vspace{.8mm}\par\noindent {\sl #1}\\
\vspace{0.2mm} }

\def\Tr{{\rm Tr}\ }
\def\tr#1{{\rm Tr}_{\tiny {#1}}\ }

\def\vev#1{\left<#1\right>}
\def\m{{\cal M}}

\def\a{{\cal A}}
\def\r{{\mathbb R}}
\def\Z{{\mathbb Z}}
\def\k{{\mathbb K}}
\def\c{{\mathbb C}}
\def\b{{\cal B}}
\def\bb{{\mathbb B}}
\def\ident{{\bf 1}}

\begin{document}

\draft
\preprint{
SU-ITP/02-39\\
hep-th/0210175}
\title{Constructing exotic D-branes with infinite matrices in type IIA string theory}
\author{Yonatan Zunger\footnote{e-mail: zunger@itp.stanford.edu}}
\address{Department of Physics\\Stanford University\\Stanford, CA 94305-4060}
\date{\today}
\begin{abstract}
We examine the set of objects which can be built in type IIA string theory by
matrix methods using an infinite number of D0-branes. In addition to 
stacks of ordinary D$p$-branes and branes in background fields, we find
exotic states which cannot be constructed by other means. These states 
exhibit strongly noncommutative geometry, (e.g., partial derivatives on them 
do not commute) and some are conjectured to have $\Z_N$-valued 
charges similar to those of the type I D-instanton. Real-valued charges are
forbidden by Dirac quantization, leading to a nontrivial relationship
between noncommutative topological invariants.
\end{abstract}
\maketitle

It is already widely known how several $p$-brane type objects may be built
out of zero-branes by matrix methods; flat membranes, compact membranes
of various geometries, and 4-branes. \cite{Matrix1} The purpose of
this paper is to determine what the most general state is that can be built
by matrix methods in IIA string theory with an infinite number of 0-branes. 
The guiding idea 
is that, with a finite number of branes, one sees what looks like a higher 
D-brane wrapping a fuzzy space. (Like the usual fuzzy sphere construction) 
However, infinite matrices allow a lot of qualitatively more exciting things
to happen; what one sees is a D-brane wrapping a generic noncommutative
space.\footnote{The idea here is that there is a pairing between topological
spaces and commutative algebras that maps $M$ onto $C_0(M)$, the algebra of
continuous functions on $M$ vanishing at infinity with pointwise addition and 
multiplication.
When we do physical calculations, typically everything is in terms of 
functions anyway, so we are already implicitly working in terms of $C_0(M)$.
A general noncommutative algebra $\a$ is then defined to be dual to a
noncommutative space. Most ideas about commutative space have a fairly
natural generalization; e.g., integrals over all of space become traces
on the algebra, and compactness of the space corresponds to the algebra
having a multiplicative unit. A ``fuzzy space'' is the space corresponding to a 
finite-dimensional (matrix) algebra $M_N$ for some $N$. (All fuzzy spaces
have the same algebra; in fact, this is the only possible finite-dimensional
algebra, part of the reason that nothing qualitatively new ever happens
at finite $N$.) Another algebra that comes up often is the irrational
rotation algebra $\a_\theta$, generated by elements $u$ and $v$ satisfying
$uv=e^{2\pi i \theta}vu$; these are the functions on the Moyal plane. 
For any algebra $\a$, $M_N(\a)$ will denote $\a\otimes M_N$, the algebra of
$N\times N$ matrices with elements in $\a$.}
All of the ordinary D-branes of type IIA can be constructed
out of (infinitely many) zero-branes in this manner, as well as D-branes in
background fields; but more interestingly, there are a large number of
exotic states which appear as bound states of infinitely many zero-branes
which cannot be constructed by other means. These states do not
necessarily have good finite-$N$ approximations; they exist only in the
strict large-$N$ limit. Finally, I can conjecture based
on these results (but not, alas, prove) that some of these exotic D-branes
have charges taking values in $\Z_N$ rather than $\Z$. (This is like the 
D-instanton in type I, which has a $\Z_2$-valued charge) The exclusion of
real-valued charges by Dirac quantization leads to an interesting relationship
between two noncommutative topological invariants.

All of these results hold at the level of the corrected non-Abelian brane
action; in fact, I was recently pleasantly surprised when a paper appeared 
\cite{KoerberSevrin} computing the corrections to the Born-Infeld action
to order $\alpha'^4F^8$, giving a result consistent with 
those given here. However, calculating beyond the Yang-Mills level requires
a great deal of mathematical machinery (pullbacks of forms onto noncommutative
spaces and so on) which is not particularly physically illuminating, 
so the body of this paper will work at the Yang-Mills level, and the
demonstration that this continues to work at all orders is left to the 
appendix. Also, nontrivial backgrounds (a metric with cycles or nonzero
background fields, for example) add significant complications, so here I
will deal only with trivial backgrounds, with occasional notes on
the generalizations which would be needed for other situations.

This paper is therefore divided as follows: I will begin by writing down
the action for an infinite number of D0-branes, which is parametrized by
an algebra $\a$ and a collection $X^\mu$ of ten covariant derivatives. (Plus
the fermions $\theta^\alpha$, of course, but these are suppressed for 
notational clarity) Using this,
I demonstrate that certain choices of $(\a,X)$ correspond to stacks of
coincident D-branes wrapping any given cycle, to D-branes in a background
B-field, and to states which correspond to neither of these. The natural
question to ask about these states is which of them are stable; the result
is that there are two topological invariants, both of which must be nonzero
for there to be a stable bound state. The first of these is an element of 
noncommutative K-theory which takes the role of the ordinary conserved
charge; it obstructs decay of the brane to closed string states by
nucleation of brane-antibrane pairs. The second is something called Hochschild
cohomology (explained below) which obstructs processes such as the classical
collapse of the fuzzy sphere down to a point under its own 
gravity.\footnote{This is the fact that the fuzzy sphere has $X=\vev{X}+A$,
where $\vev{X}$ encodes the spherical geometry and $A$ is a fluctuation, but
the minimum of the 
potential is at $X=0$ rather than $A=0$. The Hochschild cohomology class
$H^1$ essentially measures the extent to which $A$ is incapable of cancelling
$\vev{X}$.} Applied to ordinary D-branes, this leads to the usual K-theoretic 
conditions for stability; for noncommutative branes, more interesting
things may happen, such as the above-mentioned exotic states.

\section{Setting up the action and building ordinary branes}

The Yang-Mills action for a finite collection of D0-branes
is\footnote{There are two standard actions for a membrane in string theory.
The Born-Infeld action (a Yang-Mills action with corrections in $\alpha'^2 F$)
is known to describe D-branes by analysis of the S-matrix for strings 
scattering from a brane. \cite{Polchinski} The Nambu-Goto action, on the
other hand, is the ``obvious'' action for a geometric membrane, and looks
(for a 2-brane) like a gauge theory of world-volume Poisson brackets and for
a higher brane like a gauge theory of higher-order bracket-like objects. The 
former is appropriate to D-branes (which we consider here) and the latter
to membranes.\\
Both of these are gauge theories (with gauge group elements taking values in
$C(\m)\otimes M_N$ and the Poisson algebra $P(\m)$ respectively, for 
$N$ branes and world-volume $\m$) and thus have matrix approximations
which correspond to approximating the gauge algebra with some finite-dimensional
algebra. For the Born-Infeld case, we can approximate $C(\r)\otimes C(\Sigma)
\otimes M_N$ (factoring out the time component of $\m$) by $C(\r)\otimes M_k
\otimes M_N$; this is the approximation of the surface $\Sigma$ by a fuzzy
space, and generalizes the usual fuzzy sphere construction. For the
Nambu-Goto action, the algebra being approximated is the Poisson algebra;
(e.g. the Heisenberg algebra $[x,p]=1$ when the membrane is a flat plane) 
this leads to the usual matrix membrane \cite{MatrMem} and matrix 4-brane
\cite{Matr4} constructions.\\
The relationship between the two matrix constructions is the relationship 
between the Born-Infeld and Nambu-Goto actions for a brane, which is not
well understood. In the remainder of this work we will consider only D-branes
and thus the Born-Infeld action; the infinite
matrices we consider are a generalization of the fuzzy (sphere, torus,
etc.) construction.}
\be
S=\mu_0\int dt \Tr \frac{1}{4}F^2 + fermions + RR\ fields\ ,
\label{eq:oldaction}
\ee
where $F_{\mu\nu}=[X_\mu,X_\nu]$ is the field strength and 
$\mu_0=(2\pi/g)(2\pi l_s)^{-1}$. Certain aspects
of its generalization to infinite $N$ are obvious; the fields now take 
values in an algebra $\a$ more general than $M_N$, and the trace becomes a trace
on the algebra. What happens to $X$ is a bit more subtle. $X$ is not an
ordinary ($\a$-valued) field, but rather a derivation on $\a$. (i.e., a
map $\a\rightarrow\a$ that satisfies the product rule) When we say that $X$ is 
some matrix $\hat{X}$,
we really mean that as a derivation, $X(f)=[\hat{X},f]$; this satisfies the 
product rule thanks to the Jacobi identity. One obvious way to see that
$X$ is a derivation rather than a field is that in the $p$-brane action,
the components of $X$ parallel to the brane become ordinary covariant
derivatives. 

It is worth recalling a few of the properties of derivations on algebras.
(This is summarized from \cite{Palmer}, a good general textbook on the
subject) The set $\Delta(\a)$ of all derivations on $\a$ forms a Lie algebra,
since the commutator of two derivations is also a derivation. A theorem states
that every continuous automorphism of $\a$ can be written in the form
$a\rightarrow e^{i\theta\partial}a$, where $\theta$ is a real parameter
and $\partial$ is some derivation; thus this Lie algebra is the algebra
of infinitesimal automorphisms of $\a$. (The proof is very similar to the
derivation of angular momentum from rotational symmetry in quantum mechanics)
For example, when $\a=C(\r^n)$,
the ordinary partial derivatives form a basis for $\Delta(\a)$, and the
automorphisms are general coordinate transformations. When $\a=C(\r^n)\otimes
M_k$, (so the elements of the algebra are matrix-valued functions) a
basis for the derivations is $\partial_i\otimes\ident$ and $[a,\cdot]$, 
where $a$ is any matrix-valued function; the automorphisms generated by
this new generator are $U(k)$ gauge transformations. (Thus $\Delta(\a)$ is
truly the symmetry algebra of the system) 


Note that, since $X\in\Delta(\a)$, it transforms under this symmetry algebra
by commutation. (i.e., the ordinary action of a Lie algebra) Thus $X$ is
manifestly a gauge-covariant quantity, as is $F=[X,X]$. In terms of this
quantity, the action (\ref{eq:oldaction}) is also valid at infinite
$N$. Alternatively, one may absorb the integral (and thus the time coordinate)
into an additional $C(\r)$ factor in $\a$; then using the fact that the
trace on $C(\r)$ is $(2\pi l_s)^{-1}\int dt$, (the coefficient being 
for later convenience and dimensional consistency) we may write the action at 
infinite $N$ in the simpler form
\be
S=\frac{\pi}{2g}\tr{\a} F^2\ .
\label{eq:theaction}
\ee

When $\a=C(\r)\otimes M_N$, the possible derivations are $\partial_t$ 
and $[a,\cdot]$. We 
set $X_0=\partial_t+A_0$ and $X_i=\vev{X_i}+A_i$; we can always choose our 
basis so that none of the $\vev{X_i}$ contain $\partial_t$. This action
then immediately reduces to the ordinary finite-$N$ action. When 
$\a=C(\r^{p+1})\otimes M_N$, $p+1$ of the $\vev{X}$'s can be partial
derivatives, and the trace is $(2\pi l_s)^{-p-1}\int d^{p+1}\xi\ \tr{N}$,
so the action (\ref{eq:theaction}) becomes 
\be
S_{C(\r^{p+1})\otimes M_N} = \frac{2\pi}{g(2\pi l_s)^{p+1}} \int d^{p+1}\xi\ 
\tr{N} \frac{1}{4}F^2\ ,
\ee
the standard action for a stack of Dp-branes. Similarly, when $\a=C(\r)
\otimes\a_\theta$, the action looks structurally like the D2-brane action, but 
the multiplication rule for functions is now the Moyal star product;
the resulting action is that of a 2-brane in a constant background
B-field.\footnote{That is, the action of a single (commutative) D2-brane in
a background B-field is equivalent under a gauge transformation of the
2-form to the action of the given noncommutative brane without a background
field. \cite{SeiWitt}}
Note that these are all the same action, with different choices for $\a$.

Three related points have been subtly glossed over here. First, how do we 
determine which derivatives are available? Second, how is it that $A$ is 
ordinarily thought of as an $\a$-valued field, one which cannot take the 
value (e.g.) $\partial_t$, rather than a derivation, while $X$ is a generic
derivation? Third, the trace is normally defined in terms of
ordinary elements of $\a$, but equation (\ref{eq:theaction}) includes a
trace of a derivation. In the simple cases considered above, there seemed
to be a natural interpretation, but how should this be generalized?

Note that, for any algebra $\a$ and any $a\in\a$, $[a,\cdot]$ is a 
derivation. The set $\Delta_I(\a)$ of all such derivations is called the set 
of inner derivations, and it is a Lie subalgebra of $\Delta(\a)$. The first 
Hochschild cohomology class $H^1(\a)$ is defined to be the coset $\Delta(\a)/
\Delta_I(\a)$, forming a group under addition; it is a known topological 
invariant of algebras. \cite{Cohomology} For $\a=C(\r^n)$, all inner 
derivations are 
clearly zero, so $H^1$ is generated by the ordinary partial derivatives. 
For $\a=M_N$, one can show that $H^1$ is zero, so all
derivations are inner. Unfortunately, $H^1$ is in general very difficult to 
calculate; the known results are that 
\begin{enumerate}
\item When $\m$ is a manifold, $H^1(C(\m))$ is generated by the ordinary 
derivatives on $\m$, and its dimension is equal to the dimension of $\m$.
\item $H^1(\a_\theta)$ is two-dimensional, generated by the ordinary partial
derivatives with respect to $x_1$ and $x_2$. \cite{Takai}
\item When $\a$ is a von Neumann algebra (a class of algebras that includes
$M_N$, $\k$ and $\bb$\footnote{$\k$ is the algebra
of compact operators, the set of infinite matrices whose eigenvalues, listed
in descending order, go to zero; it is the algebra of infinite matrices
that appears e.g. in the matrix membrane construction. \cite{HoppeS,Harvey,Zunger}
$\bb$ is the algebra of countably infinite matrices with bounded eigenvalues.})
$H^1(\a)$ is trivial.
\item Higher Hochschild cohomology classes (see the references for details)
obey the K\"unneth formula
\be
H^n(\a\otimes\b)=\bigoplus_{p+q=n}H^p(\a)\otimes H^q(\b)\ ,
\ee
and $H^0(\a)$ is trivial for every $\a$; thus 
\be
H^1(\a\otimes\b) = H^1(\a)\oplus H^1(\b)\ .
\ee
\item Hochschild cohomology is continuous over inductive limits, i.e. 
$H^n(\lim\limits_\rightarrow \a_i)=\lim\limits_\rightarrow H^n(\a_i)$. 
This follows from
theorem 21.3.1 of \cite{Blackadar}. In particular, $H^1(\a)=0$ whenever
$\a$ is an AF-algebra. (An AF-algebra is a C*-algebra that is the limit
of a sequence of finite-dimensional algebras)
\end{enumerate}

$H^1$ is a vector space over $\a'$, the set of elements of $\a$ which commute
with all of $\a$.

This answers the first question, of how the set of allowable derivations was
determined. The answer to the second question is that the fluctuation $A$ of 
$X$ is an ``ordinary field in $\a$,'' i.e. an inner derivation in
$\Delta_I(\a)$. The physical
intuition of this is that $A$ should correspond to a bounded fluctuation,
which a partial derivative such as $\partial$ is not. (Its eigenvalues 
spread over all of $\r$) Another way
to show this is that $A$ is an inner derivation for the same reason that
on commutative space it is a function: if one begins from the definition
of a bundle in terms of transition functions, the connection is the 
Poincar\'e dual of the derivative of the log of the transition function,
and therefore an ordinary 1-form. (With function-valued coefficients) This
argument continues unchanged in the noncommutative case. \cite{Aschieri}
The third question
can now be answered by saying that the trace acting on an inner derivation
is the ordinary algebraic trace, and acting on an outer derivation
\be
\Tr \partial = 0\ .
\label{eq:fundthm}
\ee
This is simply the fundamental theorem of calculus, generalized
to noncommutative space. 

Note that adding a B-field (even a nonconstant one) to an ordinary brane
will not create all of these states, since a B-field can deform the
algebra by changing $[x_i, x_j]$ but it leaves $[\partial_i, \partial_j]=0$.
Since $\Delta(\a)$ is a Lie algebra, in general partial derivatives on 
these noncommutative spaces do not commute!

\section{Brane stability and some extraordinary branes}

We now wish to determine the criteria for a configuration specified by 
some $\a$ and $\vev{X}$ to
be stable. We cannot, in general, rely on supersymmetry alone since an
arbitrary brane configuration preserves no supercharges. Even 
in the commutative case, the BPS condition alone is not sufficient to 
describe stability, as the stability constraints often involve 
nontrivial anomaly cancellations. These are summarized in the commutative case
by K-theory, which is known to reproduce the correct phases in the
M-theory partition function \cite{DMW} and to summarize all BPS and anomaly 
conditions. \cite{MMS} We would like to generalize these considerations to 
noncommutative branes.

In our case there are two stability issues to consider: quantum-mechanical 
decay to closed string states and semi-classical collapse similar to that of 
the unsupported 
fuzzy sphere. The first of these two concerns whether or not the brane can 
decay to closed string states by nucleating brane-antibrane pairs. (Since there
are only branes in this theory, this is the only type of quantum process
which can occur) In order to do this, we need to consider stacks of
noncommutative branes and antibranes.

The most convenient language for this is the language of modules.\footnote{A
module over an algebra is like a vector space, but with $\a$-valued rather
than real-valued coefficients.} Modules over $\a$ are the noncommutative
version of bundles over a space, in the following way. On a commutative
space $\m$, to every bundle $E$ there corresponds the $C(\m)$-module
$\Gamma(E)$ of its sections. The Serre-Swan theorem \cite{Serre-Swan} states
that this is actually a duality between the set of all bundles on $\m$ and
the set of all finitely generated, projective modules over 
$C(\m)$.\footnote{A finitely generated projective module is a module $E$ 
for which there is a second module $F$ such that $E\oplus F$ is the 
trivial module $C(\m)^n$, for some finite $n$. This ``cancellation
condition'' for the modules is the module analogue of Swan's Theorem, which
states that for every vector bundle $E$ there is a bundle $F$ such that their
direct sum $E\oplus F$ is a trivial bundle.} For noncommutative spaces, we
simply define a ``bundle'' to be such a module over $\a$. Now on a 
commutative space, there is an obvious relationship (again, a pairing)
between a covariant derivative $X$ and its associated bundle; this
relationship extends to the noncommutative case \cite{Aschieri}, so we can
associate to the covariant derivative $X$ an $\a$-module which (by abuse
of notation) we also call $X$. Physically, this corresponds to the boundary
state description of a brane; the elements of $\a$ are composed of position
and momentum operators, and are thus open string operators, and the module
$X$ is a collection of kets on which they act, i.e. boundary 
states. The module notation is simply the boundary state description in
the noncommutative case.

The reason we want to use this formalism is that it makes it easy to consider
systems of multiple branes. Consider two noncommutative branes corresponding
to modules $X$ and $Y$. Then placing both branes together corresponds
to the module $X\oplus Y$, with the operators acting on them 
being $2\times 2$ matrices in $\a$. (These correspond to $XX$ and $YY$ strings
on the diagonal, and $XY$ strings on the off-diagonal) Similarly a stack
of $N$ branes would be described by a direct sum of modules, acted upon
by $M_N(\a)$. These clearly form an additive semigroup under $\oplus$.

Now imagine that we take a brane $X$ and an {\em antibrane} $Y$. Again the
module is $X\oplus Y$, acted upon by a $2\times 2$ matrix of elements of
$\a$, but now the off-diagonal elements correspond to $D\bar{D}$ strings and
thus have the opposite GSO projection. \cite{WittenK} In operator language, 
this corresponds to a graded algebra where the diagonal elements of $M_2(\a)$ 
have positive sign and the off-diagonal elements have negative sign; the 
total GSO projection is the product of the ordinary GSO projection 
and this sign factor. For notational clarity, we write $X$ as $(X,0)$ and
$Y$ as $(0,Y)$, emphasizing that there are modules and ``antimodules;'' their
sum is the graded module $(X,Y)$. 

We can now describe the quantum stability condition. We want to allow the
brane (initially described by some $(X,Y)$) to undergo nucleation of an 
arbitrary brane-antibrane pair, described by $(Z,Z)$ for some $Z$. Thus
the brane is defined only up to the equivalence relation
\be
(X,Y)\sim (X\oplus Z, Y\oplus Z)\ .
\label{eq:equivrel}
\ee
Using the fact that $X$ and $Y$ are finitely generated projective modules,
we know that there is some $Z$ for which $Y\oplus Z=0$; thus without
loss of generality, we can continue to denote our brane by $X$ alone.
The set of graded modules under addition, modulo this equivalence relation, 
is precisely the definition of the group $K_0(\a)$, the noncommutative
generalization of K-theory;\footnote{This is actually the definition of
$K_{00}(\a)$, a precursor of K-theory which agrees with $K_0$ when
the noncommutative space is ``compact,'' i.e. when $\a$ is unital. (The
distinction is similar to the requirement of compact support in the
commutative case) When $\a$ does not contain the identity, the transition
from covariant derivatives to modules does not work properly. Instead
one can define modules on the unitization $\a^+=\a\oplus\ident\otimes\c$ of
$\a$, and then mod out the group of graded modules both by the
equivalence relation (\ref{eq:equivrel}) and by the condition that $X$ be
trivial on the extra $\c$ factor. (In the commutative case, adding the
extra $C(pt.)\simeq\c$ factor is like adjoining a point at infinity, and
the extra condition is that the bundles be trivial at that point) The
group defined by this pair of equivalence relations is $K_0(\a)$.}
modulo brane nucleation, $X$ is defined only as an element of 
this group.\footnote{A technical
aside about the K-theory argument and why I (tentatively) believe it: Arguments
for the quantum stability of branes based on brane-antibrane nucleation 
are the standard way of arriving at K-theoretic charges, since they 
are both physically intuitive and quickly lead to the correct result.
(The physical correctness of K-theory as the conserved charge, at least
in the commutative case, is strongly
indicated by the nontrivial matching of phases in the IIA partition function 
obtained using K-theory to those obtained from M-theory. \cite{DMW}) 
However, one should take these arguments with a grain of salt. First, 
they are only strictly valid at weak coupling, since at strong coupling
the perturbation expansion implicit in talking about intermediate virtual
states is not guaranteed to exist. Second, in the commutative case the
brane-antibrane argument does not make clear the origin of certain 
anomaly cancellation conditions which it implies; in technical language,
one can describe commutative K-theory as a series of refinements of the
ordinary cohomological charges by means of the Atiyah-Hirzebruch Spectral
Sequence. (AHSS) The brane-antibrane anihilation condition, strictly
interpreted, only explains the first term in this sequence; higher terms
must be thought of in terms of anomaly influxes to various instanton
configurations. (The careful commutative argument based on the AHSS was
given in \cite{MMS}) Nonetheless, these brane anihilation arguments 
repeatably seem to give the ``right answer,'' namely K-theory, in the
commutative case, naturally including all higher terms in the AHSS, and
(based on the non-perturbative test of \cite{DMW}) correctly even in
strong coupling. \\
I am therefore continuing to use this argument in the noncommutative
case. We are implicitly assuming weak coupling here, but one may expect
it to continue into the strong-coupling limit as it does in the commutative
case, for similar reasons. The second argument is harder to analyze in this
case since there is no good noncommutative analogue of the AHSS approximation
to K-theory, but we should note that in the case of vanishing torsion 
(which is implicit in our assumption of no background fields) the terms
beyond the first in the commutative AHSS are automatically zero. In the
presence of torsion, moreover, the algorithm for going from a covariant
derivative to a module will be torsion-dependent, so one could reasonably
expect the correct torsion dependence to arise in such a manner. This 
belief is further reinforced by the result from category theory that
K-theory is essentially the only possible topological invariant of an
algebra which satisfies reasonable covariance properties. \cite{Cuntz}}

Two things can help clarify this: comparing it to the analogous commutative 
calculation and examining known cases. In the commutative case, one begins
with a stack of space-filling 9-branes described by a bundle $E$. (This
bundle is a collection of ten covariant derivatives, analogous to our $X$.
9-branes are used because this causes all components of $X$ to
``turn on,'' i.e. acquire a partial derivative term, so that the bundle is
over all of spacetime; in our formalism, we
used zero-branes but their infinite number allows all components to 
be turned on as well.) Stacks of branes and antibranes can be described
by pairs of bundles $(E,F)$, again with opposite GSO projections for 
brane-antibrane strings. By creating a brane-antibrane pair, we transform
this to $(E\oplus H, F\oplus H)$, and so we say that the brane state is
only defined modulo transformations of this sort. The set of bundles on
the spacetime $\m$ modulo this relationship defines the commutative K-theory 
class $K^0(\m)\simeq K_0(C(\m))$.
Then using the Sen construction \cite{Sen} to build lower branes out of higher 
branes by repeated anihilation, this result can be extended to $p<9$, giving
the same result where now $\m$ is the manifold wrapped by the brane.
\cite{WittenK}

Several things should be noted about this. First, the commutative argument
has a remarkably similar structure to our argument above, in that it involves 
pairs of branes and antibranes forming a group under addition modulo
anihilation. They differ in that the noncommutative construction begins
from 0-branes and ``builds upwards'' to form higher branes by noncommutative
methods, while the commutative argument begins from 9-branes and ``builds
downwards'' via the Sen conjecture to form lower branes. Also, the commutative
argument began from stable 9-branes, and is therefore appropriate to type 
IIB; there is an analogous argument for type IIA string theory \cite{HoravaK} 
which gives the result that $X\in K^{-1}(\m)$, where $\m$ is the spatial
part of the manifold wrapped by the brane, and $K^{-1}$ is a certain
higher K-class. (Its details are discussed in the references; we do not
need them here.)

We can (and should) compare our results to these in the cases where they
both coincide, namely $\a=C(\r^{p+1})\otimes M_N$. In order to do this,
we must first be a bit more careful about what we mean by $C(\r)$. Normally
this is used to denote $C_0(\r)$, the set of complex-valued functions of
$\r$ vanishing at infinity. (This is the algebra dual to $\r$ by the 
standard pairing of algebras and spaces) This is clearly correct for the 
spacelike
components of $\a$, since we want to consider gauge fluctuations which 
become trivial at spacelike infinity. However, it is not correct for the
time component; there is no boundary condition that forces $X$ to
go to zero at timelike infinity! Instead, the time component ought to be
$C(\r_\infty)$, the set of continuous functions on $\r\cup\left\{-\infty,
\infty\right\}$. (So that a nontrivial boundary condition may be set at
$t=\pm\infty$) This interval is homeomorphic to the closed unit interval
$[0,1]$, which is homeomorphic to a point, so $C(\r_\infty)\otimes
C_0(\r^p)\sim C(pt.)\otimes C_0(\r^p) \sim C_0(\r^p)$.
Since $K_0$ is invariant under homeomorphisms of algebras, we can perform
this substitution inside $K_0$. Also, by the stability theorem for
K-theory $K_0(\a\otimes M_N)=K_0(\a)$ for any $\a$. Thus $K_0(C(\r_\infty)
\otimes C_0(\r^p) \otimes M_N)=K_0(C_0(\r^p))$, which (by a standard result)
is $\Z$ when $p$ is even and the trivial group when $p$ is odd.\cite{AnyK}

Physically, this means that for $p$ even, the set of modules $X$ up to
brane nucleation is isomorphic to the integers, and so there is an 
integer-valued charge associated with $X$ which is not changed by nucleation. 
If one examines the calculation of $K_0(C_0(\r^p))$ in detail, (see references)
one finds that 
this charge is simply the number of branes. When $p$ is odd, however, every
$X$ is equivalent (under brane nucleation) to the trivial module, so
this $\a$ leads to unstable branes for all $X$.
This is clearly the right answer for type IIA string theory.

This concludes the discussion of quantum stability and K-theory. The second
stability issue occurs once we have `fixed' this, i.e. once $\a$ is
given and $X$ is a module (i.e. a covariant derivative) which represents
a nonzero class in $K_0(\a)$. We are now left with possible classical
fluctuations of $X$, i.e. continuous variations of $A$. The potential
stability issue is one that was alluded to before, namely the possibility
of gravitational collapse down to a point -- a situation which, while it
does include an infinite number of 0-branes, is certainly not particularly
interesting. At a simple level, this happens because (if we fix 
timelike gauge $A_0=0$) the potential is
\be
V=\tr{\a} \frac{1}{4}[X_i, X_j]^2
\ee
where the indices go over space coordinates $1\ldots 9$. The minimum of this
clearly happens when the $X$'s commute, but in general even partial derivatives
on noncommutative spaces don't commute ($\Delta(\a)$ is, after all, a Lie
algebra) and so the minimum ends up at $X=0$ rather than $A=0$. (This is exactly
what happens for the fuzzy sphere, where the $\vev{X}$ are a set of matrices
describing a sphere at finite $N$ \cite{FuzzySphere1,FuzzySphere2}) 

The obstruction to such a collapse in a trivial background is the fact
(mentioned above) that $X$ is a general derivation in $\Delta(\a)$ but $A$ 
is restricted to $\Delta_I(\a)$. Thus classical fluctuations can only move
$X$ within a fixed element of $H^1(\a)$; if a given component of $X$ has an
outer VEV, then no classical fluctuation can ever cancel it, and so $X=0$
is topologically excluded from the space of possible solutions. Clearly the
number of linearly independent components of $X$ which can be stabilized in 
this manner is
no greater than the dimension of $H^1(\a)$, so we can refer to this latter
number as the effective dimension of the brane. 

This statement is true only in the case of a trivial background. Supersymmetry 
adds extra terms to the potential which stabilize the system when $X$
wraps supersymmetric cycles. In a flat background, the only such cycles are 
infinite flat hyperplanes, and this issue does not apply; however,
even a finite-$N$ fuzzy torus can wrap a $T^2$ of the background space
and be stable. \cite{FuzzyTorusS} Similarly nontrivial Ramond-Ramond backgrounds
can stabilize configurations, as a background $C^{(3)}_{\mu\nu\lambda}$ 
does a fuzzy sphere.
Both of these can cause additional states (beyond those given here) to have
nontrivial spatial extent; the considerations discussed here are those
which determine whether a brane can be stable even in the absence of such
additional forces.

The two topological conserved charges $H^1$ and $K_0$ are compatible in the 
sense that fluctuations of $A$ within $\Delta_I(\a)$ lead to continuous 
deformations of $X$, which leave $K_0$ invariant. Thus classical fluctuations 
never change the quantum charge. 

If we examine this charge for ordinary branes, we see a very unsurprising
result. For any manifold $\m$ wrapped by a brane, $H^1(C(\m))$ is
generated by the ordinary derivatives on $\m$ (covariant derivatives
in the GR sense) and the number of stabilized dimensions is exactly the
ordinary dimension of $\m$. (Note that this happens even if the brane is
initially curved non-supersymmetrically, so that there is a nonzero potential;
the dimension of the brane nonetheless remains fixed)
Similarly for the Moyal plane, only two
components can be stabilized. For a finite number of D0-branes, however,
$H^1$ is generated by $\partial_t$ alone and there is no stabilization;
all finite-$N$ systems collapse.

Thus we find that in order for a noncommutative brane $(\a, X)$ to be
stable, $X$ must represent a nonzero class in $K_0(\a)$ and have more than
one component with a nonzero projection onto $H^1(\a)$. 
This, of course, requires that both
$K_0(\a)$ and $H^1(\a)$ be nontrivial. $K_0(\a)$ then forms the ``group
of conserved charges,'' in the sense that when multiple branes are adjoined
to one another, the charges add using the addition rule of $K_0$. 

\medskip

This leads to some interesting conjectures about the types of 
branes which may be built by this method. In order to build a brane other 
than a commutative brane or a brane in a background $B$-field, one must 
find an algebra $\a$ such that $K_0(\a)$ is nontrivial, $H^1(\a)$ is at 
least two-dimensional, and at least two derivations in $H^1(\a)$ do
not commute. ({\em cf.} the note at end of section 1)

The main obstacle to explicitly constructing such an algebra is that it is
hard to calculate $H^1$, and the list of algebras known
to have $H^1\not=0$ is unfortunately short.\footnote{The earlier list of 
results represents the present state of the art.  I attempted to extend this 
somewhat by computing $H^1$ for algebras freely generated on $N$ operators 
with a generic normal-ordering prescription. The results (after much 
computation) were that for two generators, the only algebras with nonzero 
$H^1$ are $C(\r^2)$, $C(T^2)$ and $\a_\theta$, either with or without a unit adjoined. 
(These correspond geometrically to the noncommutative torus and plane, 
respectively) For all other algebras --- which includes the Heisenberg algebra 
$[x,p]=1$, as well as various algebras with rules like $uv=\alpha v^pu^q+
\gamma$ --- $H^1$ is zero. For three generators, the problem rapidly became 
intractable even with the aid of a computer; it is an interesting open 
problem.} Getting a nontrivial $H^1$ by means of direct products alone, i.e.
by taking an algebra $\a=C(\r^{p+1})\otimes \a_0$ where $H^1(\a_0)=0$, doesn't
stabilize the brane in an interesting manner; it simply builds the unstable
algebra $\a_0$ out of $p$-branes rather than 0-branes. However, all three
of the properties specified above are believed to be generic properties of
C*-algebras; therefore we can strongly conjecture that such states exist and 
should be easily accessible given better means of computing $H^1$.

An interesting feature of these exotic branes is that not all of them have
integer-valued conserved charges. A classical theorem \cite{AnyGp} states that 
every Abelian group is $K_0$ of some algebra; this therefore raises the
possibilities of $\Z_N$ and $\r$. The former possibility is not surprising,
being a generalization of the type I D-instanton which has a $\Z_2$ charge,
also for K-theoretic reasons. \cite{WittenK} Real-valued charges, on the 
other hand, are excluded by Dirac quantization. The brane has a
monopole coupling to the $C^{(p+1)}$ Ramond-Ramond field, where $p+1$
is the effective (Hochschild) dimension; this is simply the continuum limit of 
the Myers dielectric coupling. (See appendix) Similarly it couples magnetically
to a $(d-p-3)$--form potential, so the dual monopole (if such exists;
we conjecture that this is the case but the argument may be subtle) is
a $(d-p-3)$--dimensional noncommutative object. If both objects share a
time direction but are otherwise transverse to one another, there remain
three spatial dimensions transverse to both branes. The coupling terms
are therefore $C^{(p+1)}_{0\ldots p}$ for the electric brane and 
$C^{(d-p-3)}_{0,p+1,\ldots, d-3}$, with the electric charge valued in 
the Abelian group $K_0(\a)$ and the magnetic charge in its dual group. 
If we reduce to the transverse dimensions and time, these become point 
particles coupled electrically and magnetically to a 1-form gauge field in 3+1
dimensions. We may therefore apply the ordinary Dirac quantization
argument in this case, showing that the product of the two charges must
be $2\pi$ times an integer. This can only happen if both the electric
and magnetic charges are integral, and thus if $K_0$ is either $\Z_N$ or $\Z$. 

We therefore conclude that a continuous charge is inconsistent with Dirac
quantization for noncommutative branes in the same way as it is for
commutative ones, since the branes carry a monopole coupling to the
Ramond-Ramond fields. This leads to a conjecture (no more since the
proof above is not rigorous) that so long as $H^1(\a)$ is nonzero,
$K_0(\a)$ must be $\Z$ or $\Z_N$, or phrased another way whenever $\a$ is a
simple C*-algebra and $K_0(\a)$ is continuous, then
$H^1(\a)=0$.\footnote{Simplicity is required to avoid the trivial solution
$\a=\a_0\otimes\a_1$, where the nonzero $H^1$ comes from $\a_0$ and the
large $K_0$ from $\a_1$ with $H^1(\a_1)=0$, as when an unstable noncommutative 
brane $\a_1$ is built out of $p$-branes rather than 0-branes. In this case,
we may simply factor out $\a_0$ by considering only dimensions transverse to
it, and repeat the previous argument for $\a_1$. This argument may also
fail if $H^1(\a)$ is large enough that there are fewer than three transverse
dimensions; in these cases, however, back-reaction can no longer be ignored
and so the construction of branes becomes subtle for other reasons. As a
mathematical statement about algebras (rather than a physical one about
branes), this relationship between invariants should continue to hold because
this argument could be done in any sufficiently high dimension.} This
result is consistent with known results about C*-algebras.

\bigskip

The author would like to thank Allan Adams, Simeon Hellerman and Leonard
Susskind for valuable conversations and suggestions. This work was partially
supported by the National Science Foundation under grant number PHY-9870015.

\section{Appendix: Technical details at the DBI level}

In order to make these arguments rigorous at the DBI level, we must give
a detailed prescription for the action with infinite matrices, and 
demonstrate that the identification of a stack of $p$-branes with the
algebra $C(\r^{p+1})\otimes M_N$ continues to hold. At the Yang-Mills level,
this happened because the action depended only on the full field strength
$F$ and not independently on the components parallel and transverse to
the brane. This meant that whether a given component of $X_\mu$ was a 
brane direction or not could be determined simply by whether or not its
VEV contained a partial derivative term.

For the Neveu-Schwarz term in the Born-Infeld action, a similar argument
holds, since the action may be written (following equation (26) of 
\cite{Myers}) as 
\be
S=\mu_0\int dt \Tr e^{-\phi}\sqrt{P[G+B]+ F}\ ,
\label{eq:nsaction}
\ee
where $G=\eta$ is the (flat) background metric, $B$ is the background
tensor field (zero in our case) and $\phi$ is the dilaton. Pullbacks are
defined as in the finite-matrix case. 

For the Ramond-Ramond term, however, there are subtleties due to the
Myers effect. The action for a set of D0-branes is
\be
S=\mu_0\int P[e^{\iota_X\iota_X}Ce^B]\ ,
\ee
where $C$ is the formal sum of background Ramond-Ramond fields, $B$ is
the Neveu-Schwarz tensor field, (here zero) and $\iota_X$ is the inner product 
of forms with derivatives. For finite matrices, this satisfies $\iota_X(C_\mu
dx^\mu)=X^\mu C_\mu$,
and the commutator terms coming from multiple insertions of $X$ leads
to the well-known Myers coupling of the brane to higher Ramond-Ramond
charges. (Since each insertion of an $X$ lowers the degree of the form
by one) 

To define $\iota_X(C)$ for an algebra with outer $X$, consider the case
for an ordinary manifold $\a=C(\m)$, where $X$ is a covariant derivative.
Let $d\xi^a$ be a basis for world-volume forms and $dx^\mu$ be a basis for 
spacetime forms. By linearity of the inner product of vector fields and
forms, $\iota_X(C)=\left<d\xi^a X_a, C_\mu dx^\mu\right> = 
d\xi^a \left<X_a,dx^\mu\right> C_\mu$, so we only need the latter product. 
This is defined since 1-forms are defined to be dual objects to vectors,
i.e. maps $\Delta(\a)\rightarrow\a$. For the exterior derivative of a 
function $f$ in $\a$, we can use the definition
\be
df:\ df(\partial) = \partial(f)
\ee
for all $\partial\in\Delta(\a)$. Thus
\be
\left<X_a, dx^\mu\right>=dx^\mu(X_a)=X_a(x^\mu)\ ,
\ee
where the latter is the action of $X_a$ as a derivative on the embedding
function $x^\mu$. Since $X_a=D_a$ is a covariant derivative, we can put
this together to give
\be
\iota_X(C) = d\xi^a D_a(x^\mu) C_\mu
\ee
which gives the inner product when $X_a$ is outer. When some $X$'s are 
outer and some are inner, these combine to give
\be
\iota_X(C) = d\xi^a D_a(x^\mu) C_\mu + X^i C_i
\ee
where $a$ indexes the outer $X$'s and $i$ the inner.

For $\a=C(\r^{p+1})\otimes M_N$, with $\vev{X_a}=\partial_a$ for $a=0\ldots p$,
then, the Myers coupling becomes
\bea
&&\mu_0\int_\r \tr{C(\r^p)\otimes M_N} P[\iota_X^2 C^{(3)}+\cdots ] \nn \\
&\hspace{1.ex}&=\frac{\mu_0}{(2\pi l_s)^p}\int_{\r^{p+1}} \tr{N} P[
d\xi^a D_a X^\mu d\xi^b D_b X^\nu C^{(3)}_{\mu\nu\lambda} dx^\lambda
+ \cdots] \nn \\
&\hspace{1.ex}&=\mu_p\int_{\r^{p+1}} \tr{N} d\xi^a d\xi^b d\xi^c 
D_aX^\mu D_bX^\nu D_cX^\nu C^{(3)}_{\mu\nu\lambda} + \cdots \nn \\
&\hspace{1.ex}&=\mu_p\int_{\r^{p+1}} \tr{N} P_{p+1}\left[C^{(3)}_{\mu\nu
\lambda} dx^\mu dx^\nu dx^\lambda\right] + \cdots
\eea
where $P_{p+1}$ is the pullback onto a $(p+1)$-dimensional worldvolume and
$P$ is the pullback onto the original D0-brane worldline, and the ordinary
formula for this was used in the second step. The ellipses represent similar 
terms for each higher $C^{(k)}$. The final integral keeps only terms with
$p+1$ $d\xi$'s, so the term involving $C^{(p+1)}$ contracted entirely with
outer indices gives the ordinary monopole Ramond-Ramond coupling of a 
$p$-brane. For higher forms, $p$ indices may be contracted with outer
components of $X$, and the remaining components must be contracted with
inner components, which gives the usual Myers dielectric coupling of a 
brane to higher-rank fields.

This construction continues to hold for algebras other than $C(\r^{p+1})\otimes
M_N$.  Let $\Delta^*(\a)$ be the dual space of $\Delta(\a)$; to every outer
derivation $\partial_a$ of $\Delta(\a)$ there exists a dual basis 1-form
$d\xi^a$ satisfying
\be
d\xi^a(\partial_b)=\eta^a{}_b
\ee
for some metric function $\eta^a{}_b$. The metric cannot be set equal to one
unless $\a$ is unital, but the $d\xi$'s are nonetheless paired 1-1 with the
partials since $\Delta^*(\a)\simeq\Delta(\a)$ whenever $\Delta(\a)$ is 
reflexive as a Banach space. (This is true whenever $\a$ is a C*-algebra)

Finally, the fermion terms of the action are straightforward. The basis
forms $d\xi^a$ form an $\a$-module which is a natural $O(n)$-structure 
on $\a$, where $n=\dim H^1(\a)$. The condition for this to lift to a spin
structure can be derived by exactly the same computation as for a manifold,
with ordinary cohomology now replaced by Hochschild cohomology. (i.e.,
the integral cohomology classes $H^{(1,2)}(\a, \Z_2)$ must both vanish
to guarantee orientability and spin, respectively) This allows the definition
of spin bundles, and $\Gamma$-matrices and a Dirac operator may be constructed 
out of the metric in the usual manner. The fermions themselves are $\a$-valued
fields (not derivations) and so their action may be written down in the same
manner as for non-Abelian fermions on a commutative space. Supersymmetry
is manifestly maintained by checking it in terms of components. (Of course,
any particular configuration $X_\mu$ will usually violate supersymmetry)

Corrections to the Born-Infeld action due to nontrivial commutators have 
been computed up to order $\alpha'^4 F^8$. \cite{KoerberSevrin} These 
actions affect only the Neveu-Schwarz term and are purely in terms of
the full $F$, so the previous argument continues through without change.
Thus the Born-Infeld action continues to be defined for the case of infinite
matrices, and for the particular case $\a=C(\r^{p+1})\otimes M_N$ the
action reproduces that of a stack of $N$ $p$-branes.

\end{document}